\documentclass{article}
\usepackage{dcase2021,amsmath,graphicx,url,times,booktabs, tabularx}
\usepackage{hyperref}
\usepackage{multirow}
\def\UrlBreaks{\do\/\do-}
\usepackage{verbatim}
\usepackage{balance}
\usepackage{todonotes}

\title{On the Effect of Coding Artifacts on Acoustic Scene Classification}

\name{Nagashree K. S. Rao$^{1,2}$,
       Nils Peters$^{1,2}$
       }
 \address{$^1$ University of Erlangen-Nuremberg, Erlangen, Germany \\
         $^2$ International Audio Laboratories, Erlangen, Germany\\
         \{nagashree.s.rao, nils.peters\}@fau.de\\           
  }

\begin{document}

\ninept
\maketitle

\begin{sloppy}

\begin{abstract}
Previous DCASE challenges contributed to an increase in the performance of acoustic scene classification systems.
State-of-the-art classifiers demand significant processing capabilities and memory which is challenging for resource-constrained mobile or IoT edge devices. Thus, it is more likely to deploy these models on more powerful hardware and classify audio recordings previously uploaded (or streamed) from low-power edge devices. In such scenario, the edge device may apply perceptual audio coding to reduce the transmission data rate. This paper explores the effect of perceptual audio coding on the classification performance using a DCASE 2020 challenge contribution \cite{hu2020device}. We found that classification accuracy can degrade by up to 57\% compared to classifying original (uncompressed) audio.
We further demonstrate how lossy audio compression techniques during model training can improve classification accuracy of compressed audio signals even for audio codecs and codec bitrates not included in the training process. 
\end{abstract}


\section{Introduction} \label{sec:intro}

As it can be observed in the annual DCASE challenges, classification models to understand complex acoustic soundfields are becoming increasingly robust and accurate.
The evaluation scenario indirectly presumes that scene classification is performed on the capturing device. 
However, many mobile and IoT devices are resource-constrained and may not be able to store and execute all proposed classification architectures due to limited memory and processing capabilities. Such models can be executed on more powerful personal companion devices or remote cloud servers. Figure~\ref{fig:scenarios} visualizes three scenarios where the audio capture and the scene classification are split between different devices. In such scenarios, the captured audio signals would be uploaded or streamed from the capture device to the scene classification device. The service limitations of the (often wireless) network are met by utilizing perceptual audio codecs. 
Perceptual audio coding can significantly reduce the data rate of an audio signal for storage or transmission by removing imperceivable signal components based on psychoacoustic principles \cite{painter2000perceptual}. Compared to the original audio material, perceptual audio coding does not intend to preserve the signal waveform, it rather aims to maintain a perceptually similar or even equal audio experience. The transcoding of an audio signal from one audio codec to another when passing through a network may introduce further coding artifacts. Given the maturity of perceptual audio coding, one can presume that today's perceptual audio codecs do not affect the human ability to classify acoustic scenes. But is this also true for state-of-the-art classification algorithms? In this contribution, we will study the effect of lossy perceptual coding on the model accuracy. To our knowledge, this may be the first study in the context of acoustic scene classification.


\begin{figure}[htp]
    \centering
    \includegraphics[width=0.7\linewidth]{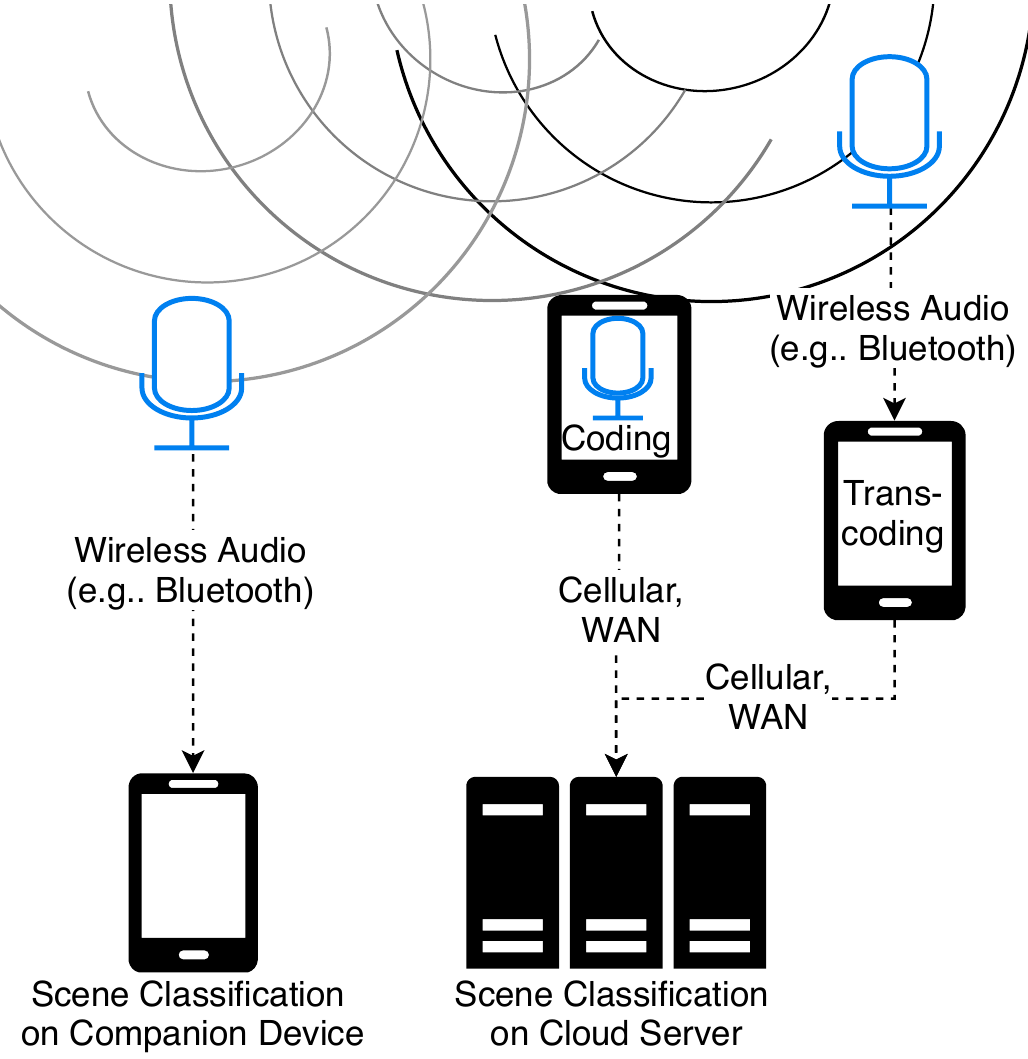}
    \caption{DCASE application scenarios where acoustic sensing and classification are decoupled and signals are coded for data transmission.}
    \label{fig:scenarios}
\end{figure}
The paper is organized as follow: We start with introducing the classification architecture we use as a basis for this study. In Section \ref{sec:expA} we present an experiment where the overall accuracy of the pre-trained reference model is evaluated as a function of differently coded audio files. Section \ref{sec:expB} reports on a series of experiments with the goal to improve the model accuracy by various training conditions. Before we conclude the paper we summarize and discuss our findings in Section \ref{sec:discussion}.


\subsection{Hu et al. DCASE 2020 Classification Model}\label{sec:Hu}
Thanks to the authors of \cite{hu2020device} who made source code and pre-trained models publicly available, we are able to study the effect of audio coding artifacts on one of the best performing models of the DCASE 2020 Scene Classification Challenge. For Task 1a, this proposal was ranked as the third best architecture, achieving a model accuracy of 76.2\% on the secret 10-class evaluation dataset\footnote{taken from challenge results at  \url{http://dcase.community/challenge2020/task-acoustic-scene-classification-results-a}}. With a total of 130 million parameters, this architecture would likely be deployed on cloud servers rather than on resource-limited edge devices.

This architecture (see Figure~\ref{fig:modelOverview}) consists of two groups of ensemble classifiers namely 3-class and 10-class. The 10-class models classify the input features (Log-mel filterbank features) to one of the ten original scene classes: airport, shopping mall, metro station, pedestrian street, public square, street traffic, tram, bus, metro, and park. The 3-class classifier is trained to predict one of the three categories: indoor, outdoor, and transportation. The final scene class is estimated by score fusion of the 3-class and the 10-class classifier.
\begin{figure}[!htp]
    \centering
    \includegraphics[width=0.91\linewidth]{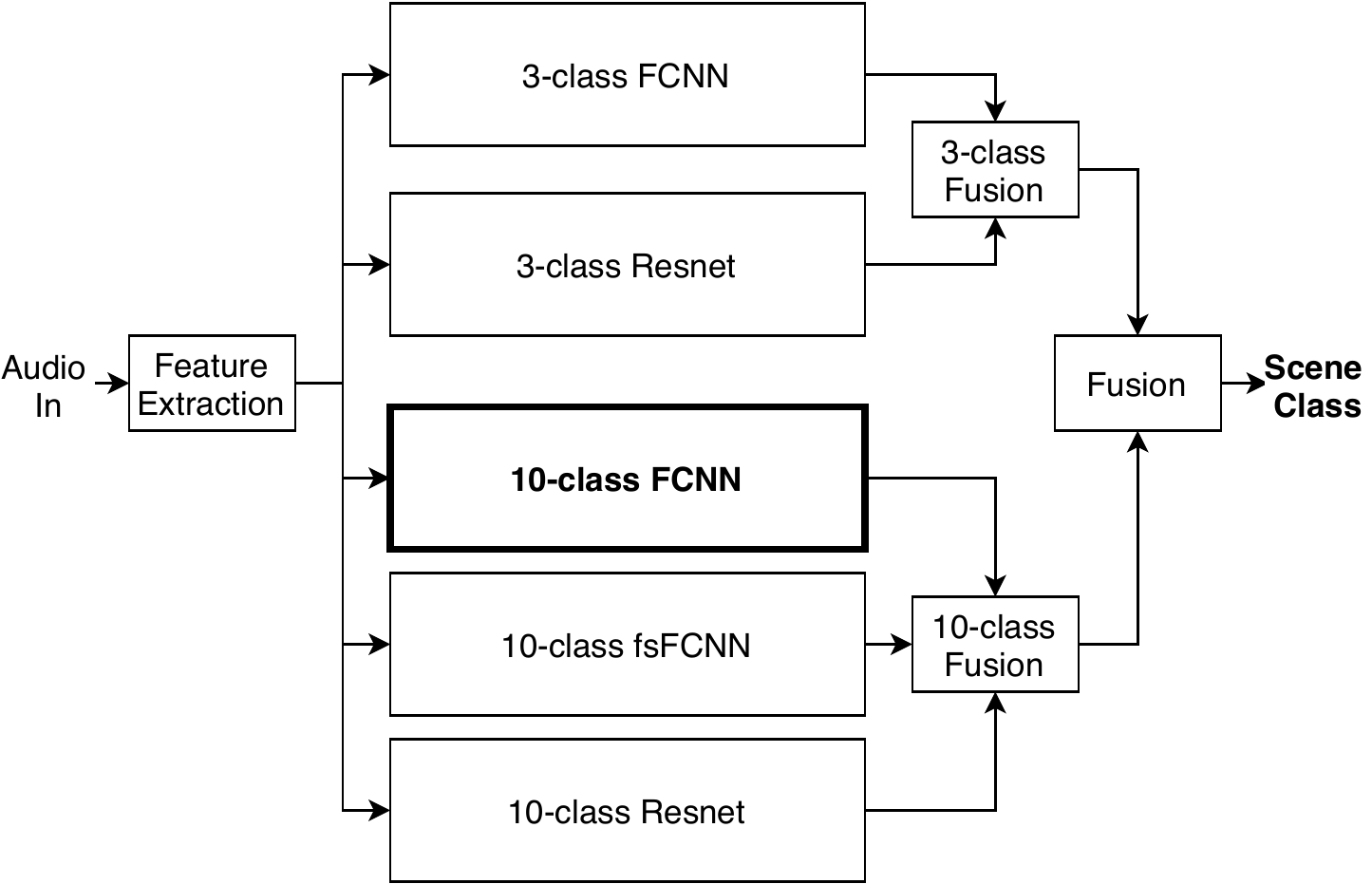}
    \caption{The scene classification ensemble architecture by \cite{hu2020device}}.  
    \label{fig:modelOverview}
\end{figure}

\section{Experiment A} \label{sec:expA}
In this experiment we want to explore whether the classification accuracy in the classification architecture described in Section \ref{sec:Hu} changes when inference is done on perceptually coded audio data rather than on original (uncompressed) audio. 

\subsection{Methodology} \label{sec:expA:methodology}

For this experiment we evaluate the complete pre-trained ensemble classifier for DCASE 2020 Task 1a by the authors of \cite{hu2020device} using the official evaluation split from the DCASE 2020 development set \cite{Heittola2020}. Figure~\ref{fig:eval} visualizes the data flow of this experiment. The evaluation split consists of 2968 monaural audio files at 44.1 kHz, each 10 seconds long. For the sake of reproducibility, all audio files are encoded using \cite{FFmpeg} with the perceptual audio codecs listed in Table \ref{tab:codecs}. The codecs are chosen because of their support in popular mobile operating systems and thus, likely to be used for compressing recorded signals prior scene classification. As model input log-mel filterbank features are extracted from the decoded audio signals. The evaluation compares the classification result with the ground truth class labels of the dataset. 
\begin{table}[!h]
\caption{Overview of perceptual audio codecs used in this paper}
    \centering
    \begin{tabular}{lcrrr}
   \hline \hline
      Codec & Reference & \multicolumn{2}{c}{Bit Rates [kbps]} \\
            &           & Exp A & Exp B\\
 \hline
 AAC (LC)& \cite{geiger2007iso} & 64 & 32, 48, 64\\
 HE-AAC (v1) & \cite{herre2008HEAAC} & 32 & 16, 32\\
 MP3 & \cite{brandenburg1999mp3} & 32, 64 & 32, 64\\
 Opus &  \cite{valin2016high} & 32 & 64 \\
 SBC & \cite{bluetooth2002b} & 64 & 64\\
         \hline \hline
    \end{tabular}
    \label{tab:codecs}
\end{table}

\begin{figure}[htp]
    \centering
    \includegraphics[width=0.48\linewidth]{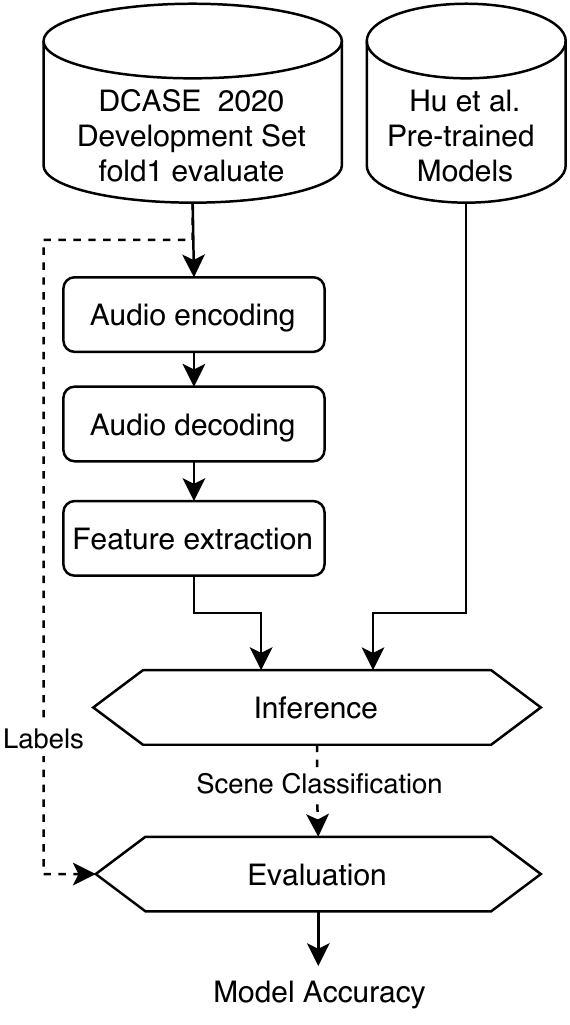}
    \caption{Exp A: Evaluation of pre-trained models with coded audio}
    \label{fig:eval}
\end{figure}

\subsection{Results}
Table \ref{tab:resultsExpA} summarizes the results of Experiment A. We found significant degradation of the overall model performance when scene classification is performed on previously compressed audio data. In our experiments, the performance of the model accuracy could drop from 0.820 (classification accuracy on the uncompressed original evaluation files) to 0.352, a decrease by 57.1\%. Not surprising, audio compression using higher bit rates (e.g., AAC at 64 kbps) tends to achieve a better model performance. 

To understand if the model accuracy changes with the perceptual audio quality due to audio coding, we computed the PEAQ objective difference grade (ODG) \cite{thiede2000peaq} between the uncompressed and the previously compressed audio files. For a comparative analysis of PEAQ and other quality metrics see \cite{torcoli2018comparing}. We computed the average ODG across all evaluation files per codec under test and compare it with the model accuracy from Table \ref{tab:resultsExpA}. As visible in Figure~\ref{fig:scatterExpA} a high ODG (i.e., little coding artifacts) results in a good model performance. However, for lower ODGs (more coding artifacts), this relationship vanishes: Two codecs that achieved an ODG around $-3.1$ result in a very different classification accuracy (0.61 vs. 0.35). Further, the model accuracy is also not a reliable predictor for PEAQ's ODG: A scene classification accuracy of about 0.65 could be achieved for codecs with an ODG between -3.1 and -2.1.
In summary, the scene classification performance correlates only weakly with PEAQ's estimated perceptual audio quality ($R^2=.44$). 
\begin{table}[h]
\caption{Exp A: Model accuracy for differently coded signals}
    \centering
    \begin{tabular}{crrr}
   \hline \hline
      Codec & Bit Rate& Model & Relative \\
            & [kbps] & Acc. & Decrease\\\hline 
      None (Original)  & 1058    & .820   & \multicolumn{1   }{c}{N/A} \\
      AAC & 64 & .741 & 9.6 \%\\ 
      MP3              & 64     & .724 & 11.7 \%\\ 
      Opus & 32 & .691 & 15.7 \%\\ 
      HE-AAC & 32  & .653 & 20.4 \%\\ 
      SBC & 64 & .632 & 22.9 \%\\ 
      MP3              & 32     & .352   & 57.1 \%\\ 
        \hline \hline
    \end{tabular}
    \label{tab:resultsExpA}
\end{table}

\begin{figure}[hbtp]
    \centering
    \includegraphics[width=0.85\linewidth]{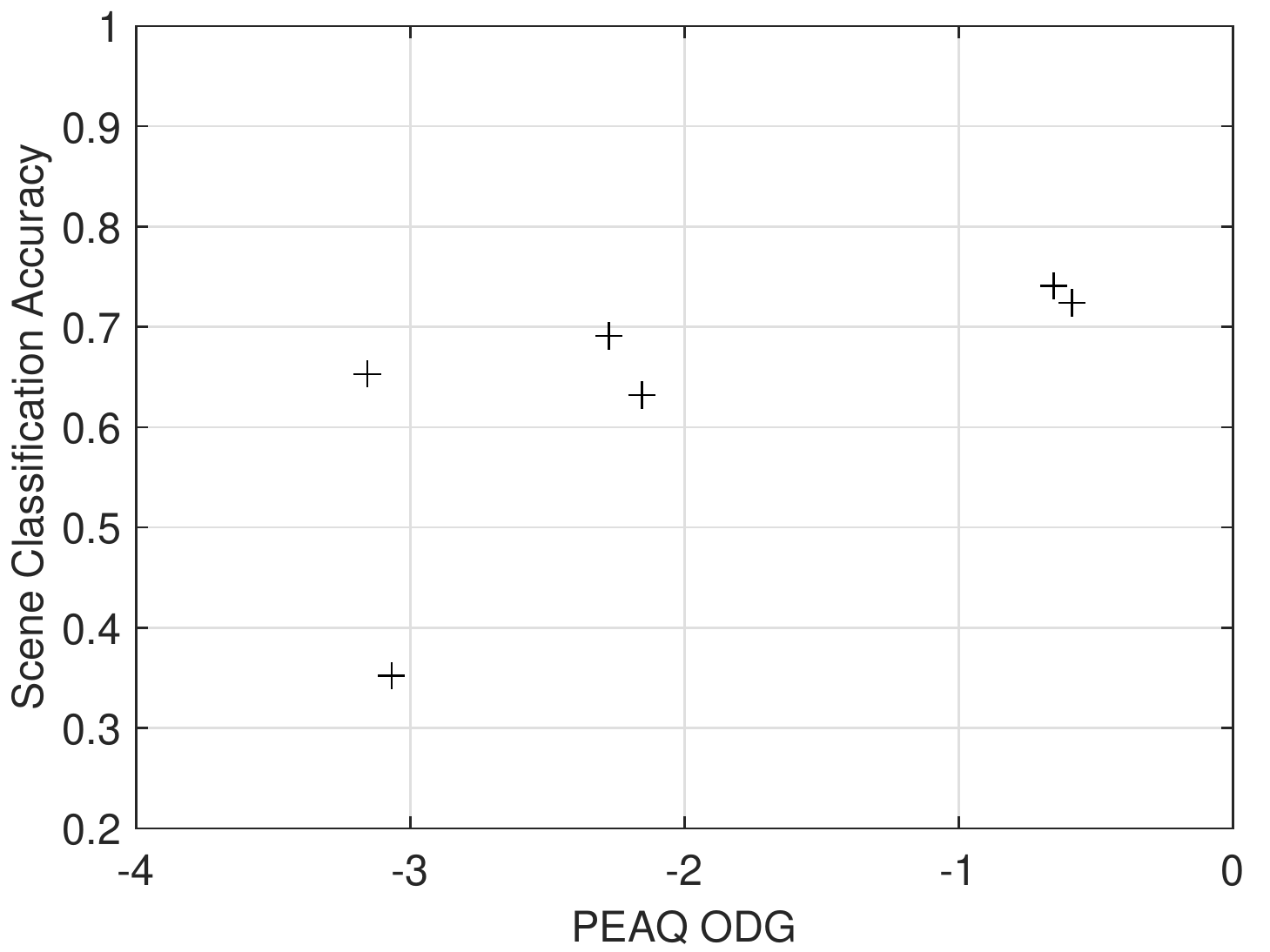}
    \caption{Experiment A: Relationship between PEAQ ODG and model accuracy for differently coded signals}
    \label{fig:scatterExpA}
\end{figure}

\section{Experiment B} \label{sec:expB}
Based on the results of Experiment A and to improve the scene classification performance of previously compressed audio data, we propose to retrain the model using additional data augmentation. In this series of experiments, we focus on the \mbox{10-class} FCNN classification sub model (see highlight in Figure~\ref{fig:modelOverview}).
These experiments also contribute to a better understand of the trade-offs between executing low-complexity classification models on the edge using uncompressed audio vs. executing complex DCASE models on a cloud server using compressed audio.

\subsection{Methodology}

The 10-class FCNN sub model is retrained with different sets of pre-compressed audio data. The workflow is shown in Figure \ref{fig:eval_expB}. All hyperparameters for the training are kept as in \cite{hu2020device} to allow for comparison across experiments. First, augmentation data are generated by converting the training dataset into different encoded audio formats and bitrates \cite{FFmpeg}. To decide which codec configurations to use for the training, we analysis the log-mel features across uncompressed and previously compressed audio data. 
The aim of this analysis was to maximize the differences in the feature data via a subset of audio codec configurations. As a result of this analysis, MP3 at 64kbps, AAC at 32kbps and HE-AAC at 16kbps and 32kbps were selected. 

The newly trained models are evaluated with the same audio content as in Experiment A in the following conditions: First, the original evaluation data are used to verify that the re-training has no degrading effects on the classification of uncoded audio data. 
Then, a subset of the evaluation conditions use codecs at bit rates that were part of the training (i.e., MP3 at 64 kbps, AAC at 32 kbps), other conditions feature the same codec used for training, but at different bit rates (i.e., MP3 at 32 kbps, AAC at 48 and 64 kbps). In the final two conditions, audio codecs that were not part of the model training are applied (i.e., Opus at 64 kbps, SBC at 64 kbps).

\begin{figure}[htp]
    \centering
    \includegraphics[width=0.8\linewidth]{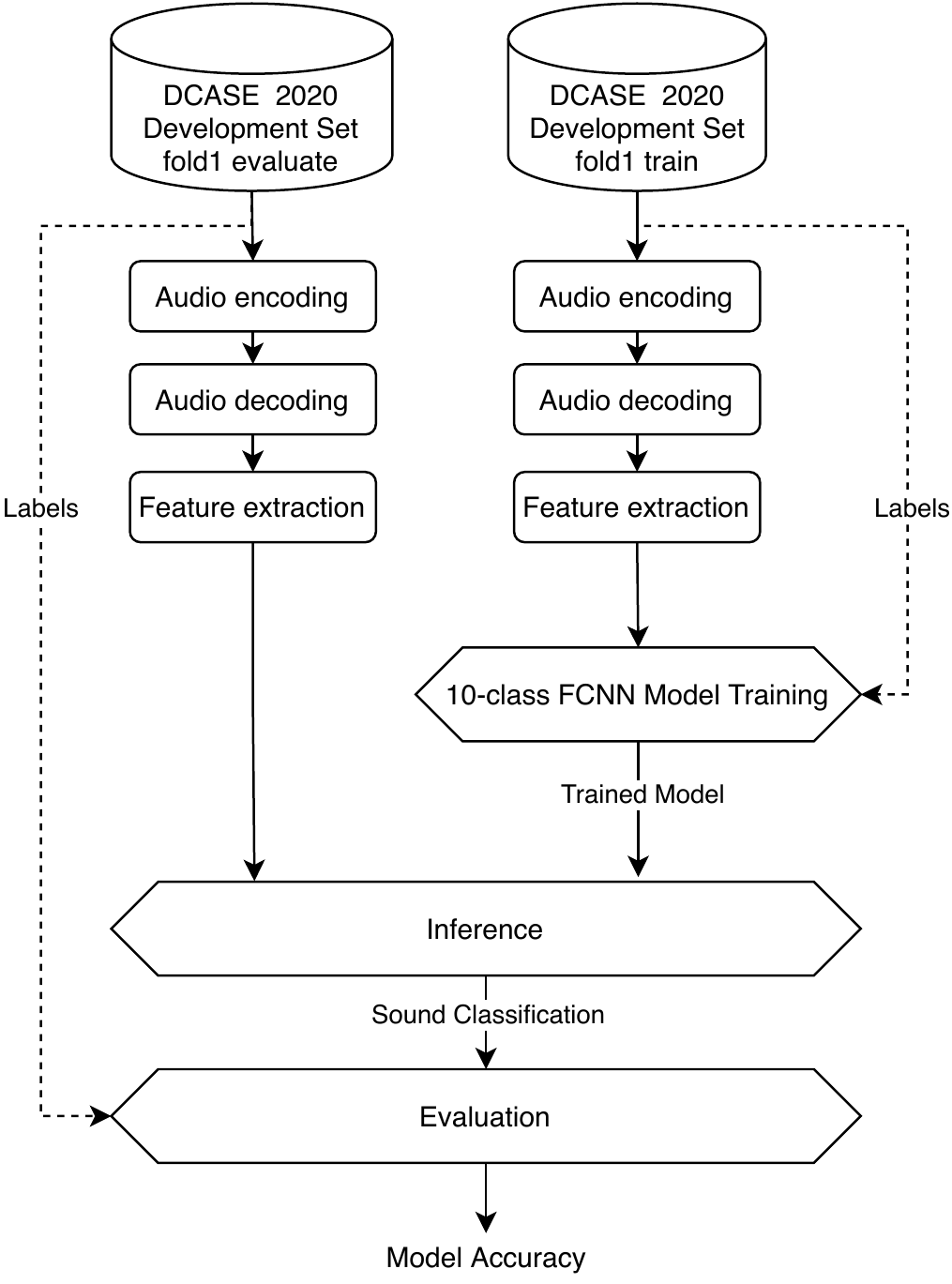}
    \caption{Exp B: Model training and evaluation with coded audio}
    \label{fig:eval_expB}
\end{figure}

\subsection{Results}

\begin{table*}[!t]
\centering
\caption{Experiment B: Model accuracy for different training conditions evaluated with different types of previously compressed signals. The subscript in the codec name indicates the bitrate in kbps}
\begin{tabular}{clr|rrrrrrrr}
\hline \hline
 \multicolumn{3}{c|}{\textbf{Training}}      & \multicolumn{7}{l}{\textbf{Codec for Evaluation}} \\
\textbf{Index} & \textbf{Condition}                        & \textbf{File count} & \textbf{None}  & \textbf{AAC\textsubscript{32}} & \textbf{MP3\textsubscript{64}} & \textbf{MP3\textsubscript{32}} & \textbf{AAC\textsubscript{48}} & \textbf{AAC\textsubscript{64}} & \textbf{Opus\textsubscript{64}} & \textbf{SBC\textsubscript{64}}  \\
\hline
1     & None (only original training data)       & 13962      & .703 & .458 & .550 & .261 & .477 & .525 & .587  & .391  \\
2     & Fully augmented data set as in \cite{hu2020device}& 121911&\textbf{.721}&.558& .615 & .301 & .573 & .622 & .638  & .555  \\
\hline
3     & 1+MP3$_{64}$                             & 27924      & .668 & .566 & .635 & .249 & .602 & .630 & .556  & .363  \\
4     & 3 + HE-AAC$_{16}$                        & 41886      & .696 & .631 & .662 & .477 & .638 & .651 & .587  & .534  \\
5     & 3 + HE-AAC$_{32}$                        & 41886      & .699 & .630 & .666 & .319 & .648 & .663 & .593  & .508  \\
6     & 4 + AAC$_{32}$                           & 55848      & .697 & \textbf{.673} & .675 & .561 & .664 & .671 & .610   & .560   \\
\hline
\multirow{2}{*}{7} & 2 + 6                                     & 163797     & .720  & .670  & \textbf{.685} & \textbf{.598} & \textbf{.685} & \textbf{.690}  & \textbf{.650}   & \textbf{.589}  \\
 & \multicolumn{2}{r|}{relative performance increase from 2 [\%]} & -0.1  & 20.1  & 11.4  & 98.7  & 19.6  & 10.9  & 1.9    & 6.1   \\
\hline
\hline
\end{tabular}
\label{tab:ExpBevalTable}
\end{table*}

The results of experiment B are summarized in the Table \ref{tab:ExpBevalTable}. From the first row in Table \ref{tab:ExpBevalTable} we can see that accuracy of the model when trained with no augmentation data and evaluated with compressed audio codecs was poor compared to the accuracy with the original audio data. When the model was trained using the data augmentation methods as proposed in \cite{hu2020device} (essentially our baseline), the accuracy when evaluated with the original files increased moderately from 0.703 to 0.721, but the accuracy was still comparatively lower when evaluated with coded signals (e.g., 0.301 for MP3\textsubscript{32}). Notably, as we start including coded audio files in the training, the performance for the codec evaluation conditions improved significantly. With inclusion of MP3 files at 64 kbps, there was decrease in the accuracy when evaluated with original files but the performance of the model improved when evaluated with coded files e.g., AAC files and MP3 files at 64 kbps. The variation of the accuracy when the model was evaluated with different codec conditions is discussed below:

    \noindent \textbf{No Coding condition:} The classification performance of uncoded audio decreased when the MP3\textsubscript{64} files were included in the training. The performance improved gradually with inclusion of HE-AAC\textsubscript{16} and AAC\textsubscript{32} and further increased when the model was trained in the final training condition (fully augmented dataset as in \cite{hu2020device}, MP3\textsubscript{64}, HE-AAC\textsubscript{16}, and AAC\textsubscript{32}). Here, the accuracy (0.72) was comparable to the baseline \cite{hu2020device}, suggesting that additional data augmentation using perceptual audio codecs does not degrade the classification of original audio signals.

    \noindent \textbf{Seen Codec With Bitrate conditions (AAC\textsubscript{32}, MP3\textsubscript{64}):}
    The performance increased with the baseline training from 0.458 to 0.558 and from 0.550 to 0.615 respectively. The accuracy further improved with the addition of MP3\textsubscript{64}, and HE-AAC.  The highest achieved accuracy for the MP3\textsubscript{64} condition was 0.685 when the model was trained with the baseline augmentation data \cite{hu2020device} as well as MP3\textsubscript{64}, AAC\textsubscript{32}, and HE-AAC\textsubscript{16} as additional training data. The highest accuracy for the AAC\textsubscript{32} condition (0.673) was obtained when the model was trained by including the HE-AAC\textsubscript{16} and AAC\textsubscript{32} training data. 
    The relative performance for the classes AAC\textsubscript{32} and MP3\textsubscript{64} improved by 20.1\% and 11.4\% respectively, as compared to the baseline performance.
    
    \noindent \textbf{Unseen Bitrate conditions (MP3\textsubscript{32}, AAC\textsubscript{48}, AAC\textsubscript{64}):}
    Here, the models were trained with MP3 and AAC but evaluated at different bitrates.
    Initially, the accuracy of these conditions was low: 0.261, 0.477, and 0.525 respectively. In case of the condition MP3\textsubscript{32} the performance of the model improved with the inclusion of HE-AAC at 16kbps and AAC at 32 kbps training data. 
    The model accuracy for both AAC evaluation conditions improved for every audio codec added to the training. The relative performance of classes  MP3\textsubscript{32}, AAC\textsubscript{48}, AAC\textsubscript{64} improved by 98.7\%, 19.6\% and 10.9\% respectively, as compared to the baseline performance.
    
    \noindent \textbf{Unseen Codec conditions (Opus\textsubscript{64}, SBC\textsubscript{64}):} Opus and SBC were not used in training. The performance of Opus\textsubscript{64} initially did not improve when augmented with different coded files, but when the model was trained with fully augmented dataset as in \cite{hu2020device} and MP3\textsubscript{64}, HE-AAC\textsubscript{16}, and AAC\textsubscript{32}, there was a slight improvement in the performance by 1.9\%. The performance of SBC\textsubscript{64} decreased with inclusion of MP3\textsubscript{64} as augmentation data but improved by the inclusion of HE-AAC\textsubscript{16} and AAC\textsubscript{64} as augmentation file. The relative classification accuracy for SBC\textsubscript{64} improved by 6.1\% compared to the baseline performance. 

In summary the final training condition resulted in an average increase in classification accuracy by 24.1\% across all 7 codec conditions. Moreover, at the final training condition, every evaluated codec condition achieved now at least 81\% of the classification accuracy as if original audio data would have been inferred. Compared to the initial training method this is a significant improvement in the robustness across codec conditions.

\section{Discussion and Future Work}\label{sec:discussion}
As a result of our experiments, we can state that perceptual compression artifacts can significantly degrade the accuracy of today's scene classification models.
Including perceptual audio coding in the data augmentation strategy:
\begin{enumerate}
    \item does neither harm nor improve model performance when classification is performed on the original audio data. 
    \item improves model accuracy when the inferred signals have been perceptually coded. 
    \item can improve model accuracy even when the inferred signals have been perceptually coded with an unseen codec or from a seen codec with an unseen bitrate configuration.
    \item seems to better harmonize the classification accuracy for input signals with unknown coding history.  
\end{enumerate}
Whereas we used the coding framework \cite{FFmpeg} for the sake of reproducibility, results may differ when other encoder implementations are used.  
Also, since we studied one recent scene classification architecture, our results may not translate perfectly to other systems. However, since many architectures use log-mel filterbank input features, we believe our findings generally hold for those architectures.  

Although our proposed data augmentation strategy is working, it generally increases the training time due to the additional training data and requires retraining of existing models. 
Thus, it would be interesting to consider alternative approaches to improve robustness against perceptual coding artifacts, e.g., hyperparameter optimization, or transfer learning \cite{mezza_feature_2020}.


\section{Conclusion}\label{sec:conclusion}
We demonstrated how lossy perceptual audio coding can degrade the scene classification accuracy of a state-of-the-art system by up to 57\% compared to uncompressed audio captures, depending on audio codec and bit rate. To increase robustness against compression artifacts, we propose a data augmentation strategy for model training that includes perceptual audio coding. We showed that such strategy can increases the accuracy when classifying perceptually coded signals even for audio codecs and/or bitrates not part of the model training. These findings are beneficial to improve scene classification robustness whenever audio signals are subject to perceptual audio coding, e.g., due to transmission or lossy data storage. 

\section{ACKNOWLEDGMENT}
The International Audio Laboratories Erlangen 
are a joint institution between the University of Erlangen-Nuremberg and Fraunhofer IIS. We would like to thank the authors of \cite{hu2020device} for making their architecture publicly available and especially Chao-Han Huck Yang and Hu Hu for their support in executing the software.

\balance 
\bibliographystyle{IEEEtran}
\bibliography{refs}

\end{sloppy}
\end{document}